\documentclass[journal]{IEEEtran}

\usepackage{colortbl}
\usepackage[pdftex]{graphicx}
\usepackage[para,online,flushleft]{threeparttable}

\ifCLASSINFOpdf

\else

\fi

\usepackage[cmex10]{amsmath}

\usepackage{amsmath, amsthm, amssymb, amsfonts, array}
\usepackage{graphicx}
\usepackage{caption}
\usepackage{subcaption}

\begin{document}
\title{Wirelessly Powered Urban Crowd Sensing over Wearables: Trading Energy for Data}
%
%
%

\author{Olga~Galinina, 
				Konstantin~Mikhaylov, 
				Kaibin~Huang, 
	      Sergey~Andreev, 
        and~Yevgeni~Koucheryavy
\thanks{O.~Galinina, S.~Andreev, and Y.~Koucheryavy are with Tampere University of Technology}
\thanks{K.~Mikhaylov is with the University of Oulu}
\thanks{K.~Huang is with the University of Hong Kong}
}

%

\maketitle
\vspace{-0.3cm}
\begin{abstract}
In this article, we put forward the mobile crowd sensing paradigm based on ubiquitous wearable devices carried by human users. The key challenge for mass user involvement into prospective urban crowd sending applications, such as monitoring of large-scale phenomena (e.g., traffic congestion and air pollution levels), is the appropriate sources of motivation. We thus advocate for the use of wireless power transfer provided in exchange for sensed data to incentivize the owners of wearables to participate in collaborative data collection. Based on this construction, we develop the novel concept of wirelessly powered crowd sensing and offer the corresponding network architecture considerations together with a systematic review of wireless charging techniques to implement it. Further, we contribute a detailed system-level feasibility study that reports on the achievable performance levels for the envisioned setup. Finally, the underlying energy--data trading mechanisms are discussed, and the work is concluded with outlining open research opportunities.  
\end{abstract}

\vspace{-0.3cm}
\vspace{-0.3cm}
\section{Crowd Sensing in Future Smart Cities}

\subsection{Towards Ubiquitous Mobile Crowd Sensing}

For the first time in the human history, over half of global population lives in cities, and this astounding number of 3.7 billion people\footnote{Refer to http://www.forbes.com/sites/danielrunde/2015/02/24/urbanization-development-opportunity/ [Accessed 10/2016]} is only expected to double by 2050. As urbanization creates increasingly larger cities, it at the same time requires more complex infrastructure to mitigate escalating social problems, including road congestion, air pollution, and public safety. This, in turn, calls for effective monitoring of community dynamics, especially since the density of urban population is becoming extremely time-dependent, as people move around. To acquire essential data for timely decision making, successful city and society management hinges on the efficiency of such monitoring during large-scale phenomena, far beyond quantifying information from a single person.

Fueled by advanced monitoring for improved maintenance of critical urban infrastructure, the vision of a smart city materializes rapidly by relying on the integration of the Internet of Things (IoT) with the information and communications technology (ICT). This unprecedented fusion impacts multiple smart city domains, such as traffic, environmental, and noise pollution assessment, as well as ambient assisted living, to help plan and optimize the management of diverse urban assets and resources. For instance, next-generation intelligent transportation systems may leverage driving speed and air quality data from a large population of commuters to reduce traffic congestion and improve air pollution monitoring. Further, this knowledge may be shared within the social sphere as well as benefit multiple healthcare and utility providers.

As advanced ICT facilitates collection of environmental, personal, and social information from a plethora of data sources across the city, emerging IoT applications promise to revolutionize monitoring and awareness, transportation and commuting, and even lifestyle and healthcare management. For our daily lives, this means faster and more reliable emergency response, controlled outbreaks of serious diseases, as well as lower risk of multi-vehicle accidents, destructive weather events, and terrorist attacks. Pursuing these important goals, big cities were historically forced to deploy large-scale proprietary sensor networks and then rely on legacy wireless sensor networking (WSN) technologies. However, the conventional WSN solutions have failed to proliferate widely in the real world, primarily due to their high installation and maintenance costs as well as limited coverage and scalability.

Overcoming the limitations of commercial WSN deployments, a novel knowledge discovery paradigm of mobile crowd sensing (MCS) has recently emerged to extract information from a multitude of user-paired devices with limited sensing capabilities. Generally, leveraging the power of citizens for massive sensing may assume either of the two forms~\cite{6069707}, opportunistic (autonomous data collection without direct involvement of participants) or participatory (with active interaction of humans, who decide to contribute data). An evolution of participatory sensing, MCS engages individuals with their companion devices into sharing contextual data and extracting relevant information to collectively measure and map large-scale phenomena of common interest~\cite{6815273}. To this end, MCS employs various sensors built into user devices to gather relevant information with its further aggregation in the cloud for data fusion and intelligence mining.

With ubiquitous availability of companion user devices, such as mobile phones, smart vehicles, and wearables, MCS has the potential to overcome the constraints of past WSN deployments, including their limited space-time coverage. The available information on users and their surrounding environment includes but is not limited to location, acceleration, temperature, noise level, traffic conditions, pollution, etc. By efficiently combining personal and collective data, future MCS applications may offer rich information on urban dynamics by generating knowledge about e.g., safety-related accidents~\cite{7289253} and thus provide dynamic situational awareness.

\subsection{Urban Crowd Sensing over Wearables}

The emerging vision of urban crowd sensing inherently relies on people who are, in essence, walking sensor networks. Beyond their sensor-rich handheld mobile equipment, the increasingly widespread wearable devices are becoming to play a major role in collaborative data collection. Today's consumer wearables already have a variety of sensing, computing, and communication capabilities, as they feature a host of embedded sensors for determining information on location (GPS and other wireless interfaces), positioning (digital compass, gyroscope), activity (accelerometer), noise (microphone), environment (temperature, light, humidity, pollution sensors), health (heart rate, blood pressure, stress level sensors), human social relationships (texting, Facebook/Twitter profiles), and even emotions (camera). Powered by further miniaturization of sensing components~\cite{201695}, novel classes of wearables promise to transform crowd sensing into a new global utility.

However, before urban crowd sensing over wearables can truly take off, many research challenges need to be resolved on the way to its mass penetration. Today, resource constraints of wearables in terms of their energy (battery lifetime limitations), bandwidth (link capacity and data transmission latency), and computation (sensing and processing costs) constitute the major user adoption barriers. Further, wearables are person-centric in the sense that they collect and communicate information only about their specific wearer, which may raise privacy concerns that need to be addressed with e.g., adaptive obfuscation mechanisms~\cite{6636891}. Finally, small numbers of involved participants may compromise the efficiency of a crowd sensing service and thus scalable solutions are needed for incentivizing user involvement into data sharing.


Fortunately, there is a recent innovation that holds a promise to resolve the major impediments to massive crowd sensing based on wearable devices, which is to equip them with energy harvesting capabilities~\cite{7462483}. Beyond the state-of-the-art approaches to minimize power consumption~\cite{2014149}, energy harvesting may effectively replenish the charge levels of small-scale and battery-powered wearables. Coupling the dedicated wireless charging by the surrounding wireless network infrastructure with energy-efficient data transmission protocols~\cite{Xiao2016} may supply the constrained wireless-powered wearables that run sensitive tasks with predictable amounts of energy. To advance this thinking further, we envision that energy transfer and harvesting technologies may open the door to genuinely incentive-aware crowd sensing applications, where sensed data is provided by the user to the cloud in exchange for wireless charging service.

\begin{figure}[!ht]
\centering
\includegraphics[width=1\columnwidth]{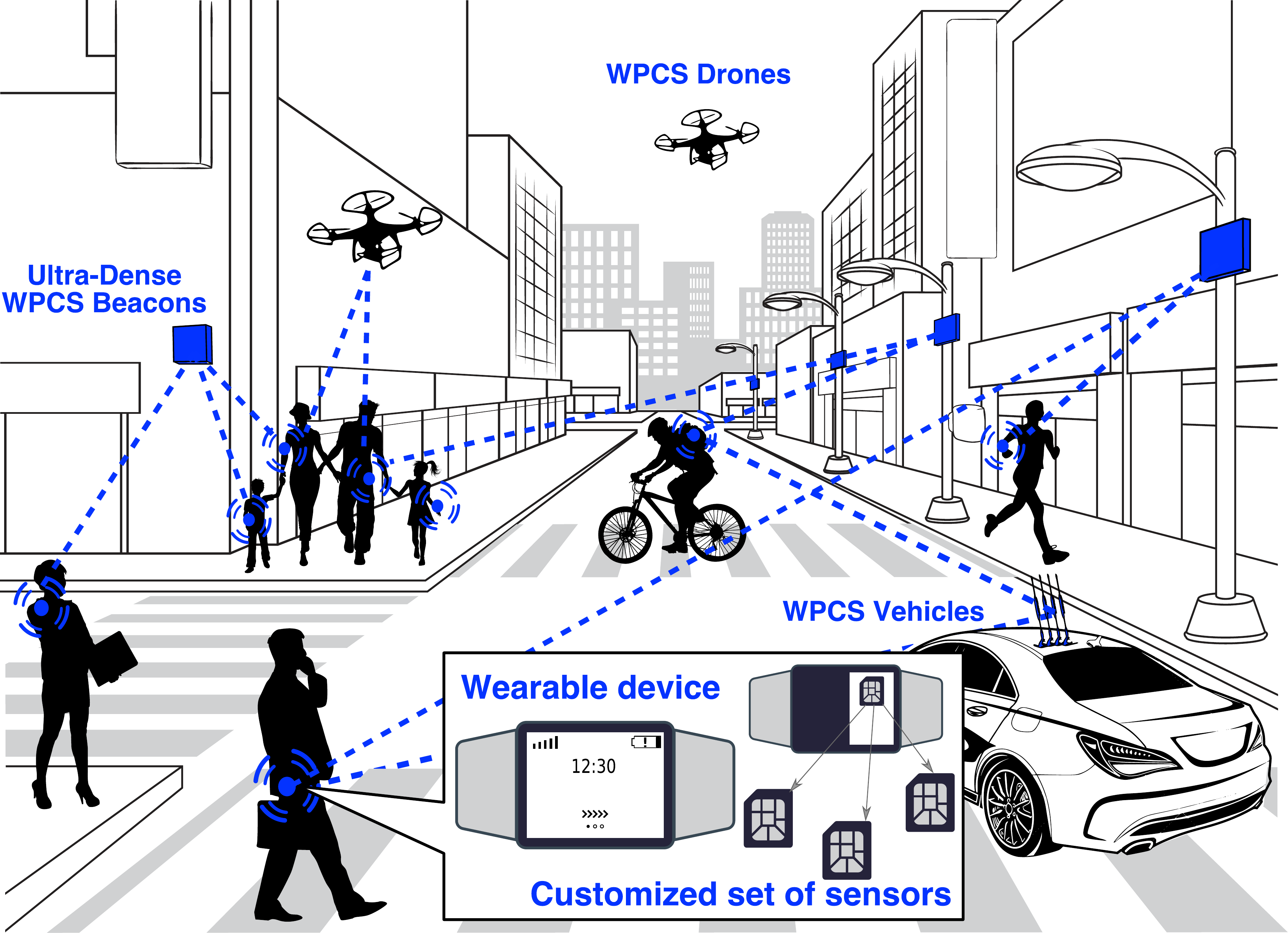}
\caption{Our vision: wirelessly powered urban crowd sensing}
\label{fig:1}
\vspace{-0.7cm}
\end{figure}

As a result of incentivized human involvement, future crowd sensing applications may engage a critical mass of people to contribute their accurate and relevant data. As an example, commuters can supply abundant information on their daily routes~\cite{6341381} to better monitor traffic conditions in a smart city. However, the opportunistic properties of unconstrained human mobility when sensing and transmitting information have seldom been addressed in the past research literature on MCS. Indeed, given that large-scale movement of users is uncertain and skewed, it needs to be captured explicitly as people travel and observe phenomena of common interest (traffic/road conditions, noise and air pollution, etc.). At the same time, inherent mobility of people with wearables offers unprecedented benefits for improved sensing coverage and space-time data collection, since humans naturally congregate in the areas where crowd sensing is more valuable.

In what follows, we first offer our vision of a novel wirelessly powered urban crowd sensing system and then conduct its thorough system-level performance assessment augmented by a review of energy--data trading mechanisms.

\vspace{-0.3cm}
\section{Wirelessly Powered Crowd Sensing System}

\begin{figure}[!ht]
\centering
\includegraphics[width=1\columnwidth]{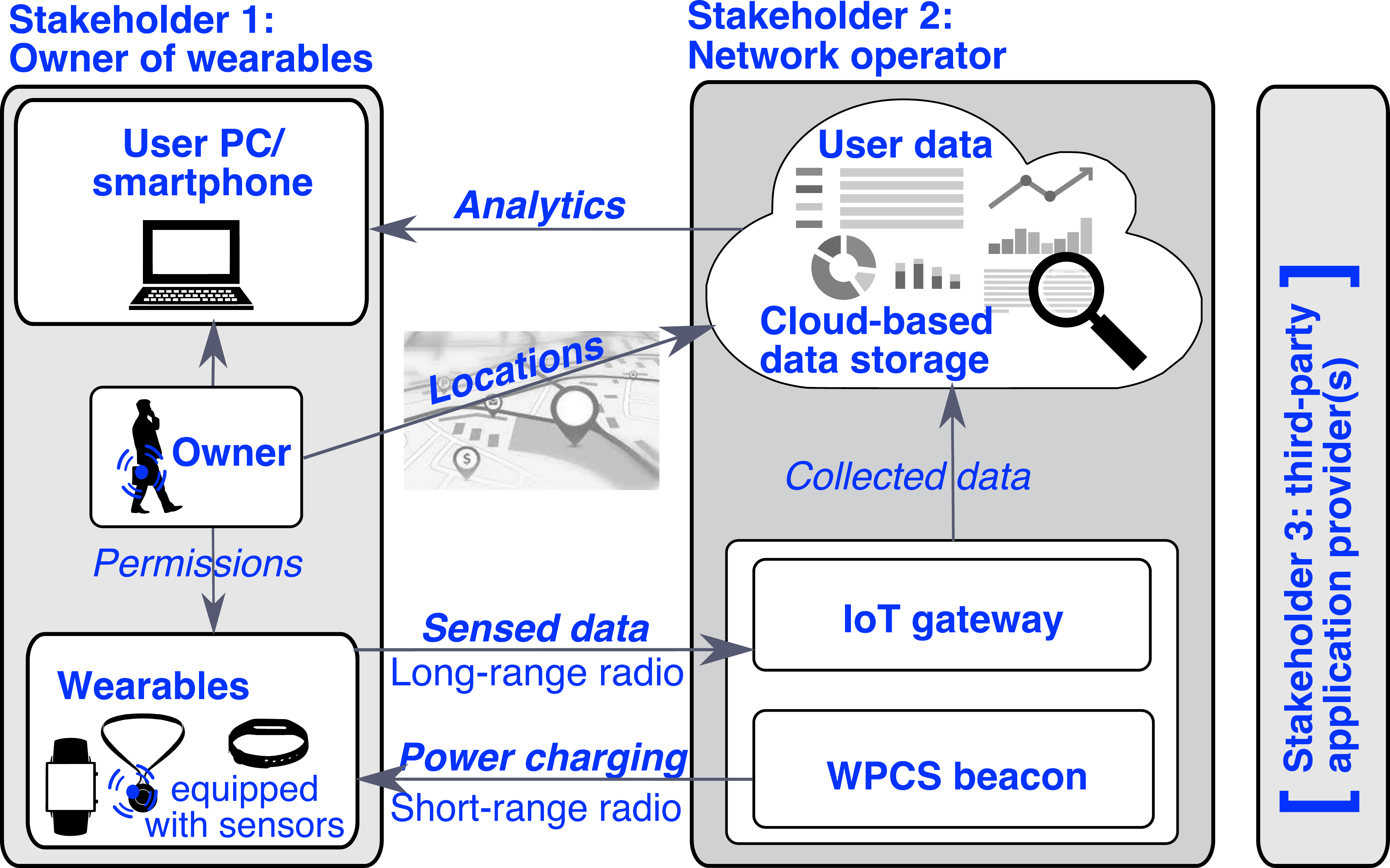}
\caption{High-level architecture of envisioned WPCS system}
\label{fig:2}
\vspace{-0.7cm}
\end{figure}

Our envisioned wirelessly powered crowd sensing (WPCS) system and its high-level architecture are displayed in Fig.~\ref{fig:2}. It comprises two major components: (i) the sensing-enabled wearables carried by people and (ii) the operator infrastructure that is deployed in the surrounding environment. Further, we detail our vision on the key components and their functionality in our proposed WPCS system, as well as discuss on the roles and motivations of the involved stakeholders. 

\subsection{Wearable User Devices}

The user-side component of the proposed WPCS system in represented by a personal network of wearable devices (e.g., smart watches, wristbands, smart clothes), potentially including other carried devices (an MP3 player or a headset, a smartphone, etc.). Generally, every wearable device includes a number of sensors and a wireless communication module that is responsible for (i) transmission of sensed data and (ii) collection of wireless power for improved device operation. Practically, both of these functions can either rely on a single component (e.g., a type of backscatter design) or be distributed across several components (e.g., out-of-band energy transfer)\footnote{More details on the energy transfer possibilities are offered in Section~III.}. In more detail, a wearable device may include a processing system, an energy buffer (i.e., a battery), a memory unit for storing data, a user interface, and other charging options.


In practice, some of the wearables (or sensors of a particular wearable device) may be temporarily disabled by the owner if not needed immediately (e.g., to save energy). We propose that with due user permission and subject to an appropriate incentive these wearables (sensors) may be made active to participate in massive crowd sensing. Hence, together with basic data collection and forwarding capabilities, the envisioned WPCS system design needs to implement two important functions. The first is the identification of the source of sensed data to facilitate the data quality assessment as well as to allow the owner of a wearable device to collect reward for sharing information~\cite{5657730}. Here, a viable solution has to also maintain adequate privacy and anonymity levels. Note that some of today's radio protocols already have sufficient security and identification mechanisms available, but for many other (and especially those based on backscatter principles) these may need to be developed. 

The second important feature is the need of localization and tracking mechanisms for wearables in order to properly map their data as well as to manage direct power transfer to them, if desired. Even though it is feasible to assume that most of the time a wearable device resides close to its owner -- and thus its localization is identical to the localization of the wearer (e.g., with wireless positioning or machine vision solutions) -- new methods for tracking individual wearables are crucial as well e.g., to enable directional power transfer. 

\vspace{-0.3cm}
\subsection{Mobile Operator Infrastructure}

The mobile operator infrastructure comprises \textit{four major components}. Note that not all of these components need to be owned by an operator -- some of the required services can be provided by the third parties. The first component of the envisioned architecture is the \textit{localization system} that enables close to real-time tracking of the positions of the wearables as well as collection of meta data about their capabilities (e.g., available sensors and their characteristics). The said system can be built on a singe or multiple combined localization technologies. The second necessary component is the \textit{incentivization mechanisms} to engage the owners of wearables into sharing their collected data for crowd sensing applications. Even though alternative schemes can be used in this context, in this work we advocate for the provision of wireless energy as an attractive incentive. 

As a result, an operator may need to deploy an infrastructure of wireless charging stations (shown in Fig.~\ref{fig:1}). We consider the following possibilities of how the energy can be delivered to a wearable device: (i) use of a charging terminal (the user has to approach a dedicated terminal and charge its devices with e.g., inductive coupling mechanisms); (ii) free-walk charging from a stationary power beacon (once the user approaches the designated power beacon, the latter begins to emit radio waves for a wearable device to replenish its energy); (iii) charging from uncontrolled mobile beacons (the difference with the previous case is in that the beacon is deployed on a mobile object (e.g., a vehicle), with its own mobility pattern that is out of control for the charging system); (iv) charging from controlled mobile beacons (similar to the previous case, but the mobility of the beacon carrier (e.g., a drone) can be controlled by the charging system). A particular WPCS system implementation may employ one or several of the above options for providing wireless energy to wearables. 

An important component of our architecture is the \textit{cloud-based data storage and processing system}, which aggregates all of the data from the users. This information can be delivered by various radio interfaces. Prior to leveraging the sensed data, their validity and authenticity (often referred to as data quality) have to be confirmed. We expect that thus collected data can be made available to both the owners of wearables that have contributed this information (hence providing another, indirect source of motivation for them to participate) as well as to the third-party services (after respective aggregation and de-personification). The final component of our WPCS system is a set of dedicated \textit{cross-layer algorithms} that track wearables as well as control the flows of data and energy.

\vspace{-0.3cm}
\subsection{Involved Parties and Their Interests}

All in all, there are three types of stakeholders that coexist in the outlined ecosystem (see Fig.~\ref{fig:2}). The first is the owners of wearable devices that host various sensors. The motivation for them to participate in the envisioned WPCS system operation is the following. The cloud storage capability provided by the system enables easy access to personal data from multiple user wearables as well as helps reduce the computation load and related energy expenditures for data analysis due to the use of cloud services. Further, some of the sensed data might not make much sense alone, but may dramatically increase in its value if aggregated. The provision of user access to the aggregated data of their wearables in exchange for sharing this information is an indirect motivation mechanism. On top of that, for a wide variety of the non-personal information (e.g., environmental parameters), the redundancy of data coming from multiple spatially co-allocated sensors can be high. In this case, the system may enforce control over sensors by reducing their duty cycle or even switching some of them off to save energy. 

Most importantly, the WPCS system may employ direct incentivization mechanism (such as wireless charging) to better motivate the users to share their data. Here, privacy-centric schemes are needed to define who and under what conditions can use such data generated by personal wearables. Depending on the implementation, the interactions within the system can be built either over (i) short- and/or real-time relations (e.g., near-immediate data--energy exchange) or on (ii) long-term engagement (e.g., subscription-based contracts). Note that real-time operation requires on-the-fly localization and discovery mechanisms. Another aspect is related to how different devices are handled, that is, whether the payoff is provided to particular device(s) or to the user network as a whole. We consider the latter option to be more advantageous, albeit it introduces challenges in mediating between the needs of individual wearables. 

The second party involved into our WPCS system is responsible for its management and is referred to as the operator. Its target is collection and aggregation of the sensory data that could then be monetized to cover expenses and generate revenues. The two approaches to reach this goal are (i) to charge the owners of wearables with a service fee or (ii) to monetize access to the collected data via the third parties (e.g., other service providers). While the former approach is simpler in terms of the corresponding system architecture, it has two major challenges. One is identifying the sources of motivation for the user to purchase such a service for a personal network of wearables. If resolved, another one is rooted in the need to provide acceptable service quality guarantees to reduce customer churn and keep people engaged. As these challenges are non-trivial, we consider the alternative approach.

Accordingly, we assume the presence of the third-party services as another (possible) type of stakeholders in our WPCS system. These are interested in collecting crowd sensed information and are ready to pay for it. Examples include governmental institutions and smart city administration (willing to analyze environmental conditions and commuter behavior), marketing and commercial agencies (targeting to understand the customer behavior in shopping malls or on the streets), and public transportation companies (assessing their traffic flows). Therefore, the operator of the proposed system needs to first aggregate and pre-process the big sensed data (as well as employ the discussed incentivization mechanisms to better motivate the users), and then provide this information to a third party. Apparently, the operator also has to negotiate the ownership rights for thus collected data (or the aggregated/processed data), but we leave this aspect out of the scope here. 

\vspace{-0.3cm}
\section{Power Transfer Options for Crowd Sensing}

In this section, we review and discuss the available wireless power transfer (WPT) options for our WPCS system.

\vspace{-0.3cm}
\subsection{WPT Techniques}

\subsubsection{Energy Beamforming} The two key technologies for WPT are resonance inductive coupling and microwave power transfer. The former is a near-field non-radiative technology suitable for short-range (less than a meter) static WPT. In contrast, the latter is much more versatile and supports longer ranges (e.g., up to tens of meters), including the cases of high mobility and multi-user WPT among other features. For this reason, we focus solely on this technology for the envisioned WPCS system. 

The main hurdle behind efficient microwave power transfer is high propagation loss. Energy beamforming, which utilizes an antenna array to steer the radiation power in a desired direction, is a basic technique for mitigating such loss. The WPT efficiency, defined as the ratio between the receive and the transmit power levels, increases linearly with the number of transmit antennas. The latest advances in the massive MIMO technology can provision WPCS charging stations with large-scale arrays, which host hundreds of antennas able to create ultra-sharp beams that suppress propagation loss and achieve high power transfer efficiency. Moreover, with recent breakthroughs in millimeter-wave communication, antenna sizes and spacing can be reduced further down to the scale of millimeters, thus dramatically shrinking the form factors of large-scale arrays. Consequently, ultra-compact WPCS stations capable of sharp beamforming could be deployed ubiquitously within an urban environment e.g., on the walls and lamp posts. 

Energy beamforming for a point-to-point static WPT is relatively simple and involves steering a fixed beam pointing at the intended location. In the context of crowd sensing however, such beamforming has to adapt and track the time-varying locations of mobile wearables. This may require the latter to periodically transmit pilot sequences to the WPCS charging stations that estimate their locations. In the process of WPT, a station needs to monitor the WPT channel gain based on the feedback from a wearable device to maintain a certain power-transfer efficiency. In particular, it may be desirable to pause WPT whenever the channel loses the line-of-sight (LoS) condition due to blockage, to avoid power waste or for the sake of safety. Prior to energy beamforming, it is necessary for the station to negotiate the power-and-data exchange with wearables, schedule a sub-set of them for WPT depending on their data availability, and then determine the corresponding power levels. Charging several devices simultaneously is possible by steering multiple beams (possibly, with different powers) while using a single array. 

\subsubsection{Cooperative WPT} For more efficient WPT, it is essential to have the LoS condition between a charging station and the target wearable device. This may not be always feasible in urban environments where the proposed crowd sensing system is to be deployed. The paths from stations to wearables may be frequently occluded by objects such as buildings, trees, and human bodies. However, if multiple WPCS charging stations serve a single device, the chances of establishing the LoS link grow exponentially with the number of collaborating stations, which is named cooperative WPT. In practice, cooperation essentially means that the stations control the phase shifts of their transmitted radio waves such that these are combined constructively at the target device. This effectively creates a virtual distributed antenna array comprising all of the antennas at the cooperating stations, which forms a virtual beam towards the charged device. 

The coordination overhead for cooperative WPT remains much lower than that for cooperative data transmission in e.g., cellular networks, where base stations need to exchange both channel information and data to transmit multiple data streams as well as mitigate mutual interference. Similarly to cooperative WPT, multiple base stations can also cooperate to collect data from wearables: the identical transmissions received from different devices may be combined coherently in the cloud to enhance the total received signal power. In the presence of several wearables, MIMO techniques can further be applied to decouple and detect multiple data streams.

\subsubsection{Simultaneous Wireless Information-and-Power Transfer} Having Internet connectivity in addition to WPT capabilities, the WPCS charging stations can further incentivize wearables to participate in massive crowd sensing applications by acting as Internet access points and delivering to wearables the information of their interest. Alternatively, WPCS stations can operate as relays and assist in communication with e.g., cellular networks. In these cases, a station performs simultaneous wireless information-and-power transfer (SWIPT) to a wearable device~\cite{7120022} by transmitting a modulated carrier wave. Note that said wave can be unmodulated in the WPT-only case. 

Then, the wearables employ a rectenna (an integrated antenna and RF energy harvesting module) to receive power together with a separate radio antenna unit to retrieve meaningful information from the same wave. Alternatively, the harvester and the receiver can share a single antenna followed by a signal splitter that divides the received RF signal into harvesting and information detection components. From the WPT perspective, modulation has mild effects on the energy harvesting efficiency (i.e., the wireless charging efficiency). However, from the information-transfer perspective, SWIPT generates interference to nearby communication links on the same band. In contrast, unmodulated waves for WPT-only case are single-tone signals and can be easily canceled out by information receivers. 

\vspace{-0.3cm}
\subsection{WPCS Charging Stations}

The WPCS charging stations can generally be deployed at fixed locations or mounted on moving vehicles and drones. The corresponding options are named here the WPCS beacons, vehicles, and drones, respectively. Their design and implementation principles are discussed individually as follows.

\subsubsection{Ultra-Dense WPCS Beacons} Together with energy beamforming, reducing the propagation distance is another way of improving the WPT efficiency. Highly effective WPT for charging wearables requires the distances of not larger than tens of meters~\cite{7462482}. Hence, ultra-dense WPCS beacons may need to be deployed, which then makes it possible to serve the massive numbers of wearables across a smart city. A practical approach to materialize the ultra-dense WPCS beacon deployments is to leverage the corresponding small-cell base stations to be available in next-generation cellular infrastructure that can be co-located with the WPCS beacons. In addition, dedicated WPCS beacons may be installed in locations where wearables are not within the WPT ranges of small-cell base stations.

Upgrading the protocols that run on top of ultra-dense small-cell deployments to offer the WPCS services can however incur substantial extra costs. The latter can be reduced in next-generation cloud radio access networks that feature rich virtualization mechanisms. This is because direct upgrading may involve software changes at data centers. On the other hand, dedicated WPCS beacons have a much lower complexity compared to small-cell base stations as they do not need to support communication. Therefore, they merely require the Internet access for uploading their collected data in contrast to more sophisticated backhaul networks. Furthermore, the algorithms for WPCS have a much lighter complexity than those for communication. Consequently, dedicated WPCS beacons can be made more compact than the cellular base stations, and thus allow for an ultra-dense deployment at low costs.

\subsubsection{WPCS Vehicles} Making WPCS stations mobile by mounting them on moving vehicles can compensate for their insufficient densities in certain areas. Moreover, when approaching the intended wearable device(s), the WPCS vehicles can further reduce the WPT distances, thereby improving the efficiency of charging. In practice, each vehicle might be assigned a fixed route along which the target wearables are located. Such a vehicle may travel along its route periodically to recharge the wearables and collect their sensed data by following the approach of~\cite{6153401} for mobile WPT. Specifically, the vehicle can either upload the aggregated data into the cloud at the end of its trip (via a wired Internet access if the information is delay-tolerant), or otherwise the upload can be made real-time by utilizing the wireless broadband access. With the rapid advancement in smart navigation, the WPCS vehicles can ultimately be made autonomous. 

One key future challenge in deploying mobile WPCS stations is the routing of WPCS vehicles. First, various types of wearables distributed across a smart city need to be mapped. Given this map, the subject areas that have to be covered by the WPCS vehicles should be identified mindful of other types of charging stations, such as static WPCS beacons and WPCS drones. The routes for vehicles are then optimized to minimize the travel distances and periods under various constraints, such as the WPCS mission, the device lifetime, and the traffic congestion cycles. For the case of SWIPT discussed earlier, the WPCS vehicles may further be equipped with a storage module for caching the content useful for the served wearables. 

\subsubsection{WPCS Drones} The WPCS drones can deliver the wireless charging services to regions where it is difficult to deploy static beacons or access with vehicles, such as parks and lakes. The drones can be designed to fly autonomously as well as be powered by solar energy. The design issues for WPCS drones are similar to those for WPCS vehicles and require careful mapping of wearables as well as optimizing the routes and travel periods. The drones can be wirelessly connected to the cellular network infrastructure for the purposes of data transfer. 

One important challenge for deploying the WPCS drones is in safety constraints. For example, the current FAA guidelines only allow for a government public safety agency to operate an unmanned aircraft with the weight of 4.4 pounds or less (within the LoS to the operator) and under 400 feet above the ground. However, the industry expects these policies to be relaxed in the future. Many companies, including Amazon and Facebook, are actively developing drone-based services such as goods delivery. This makes urban drone-based WPCS services a viable but futuristic solution. In addition, WPCS drones are expected to be deployed in sparsely populated areas where the safety concerns are less pressing. 

\begin{table}[!ht]
\setcounter{footnote}{0}
\setlength{\tabcolsep}{2.5 pt}
\renewcommand{\arraystretch}{1.3}
\caption{WPCS system composition and its main parameters}
\label{sim_parameters}
\centering
{\footnotesize
\begin{tabular}{|c|c|}
\hline
\bfseries Parameter & \bfseries Value\\
\hline
WPCS beacon height/User device height&3m/1.2m  \\
User height/Body diameter&1.7m/0.4m \\
\hline
WPT frequency&915MHz \\
Transmit power/Transmitter gain&1W/0dBm  \\
Sensitivity&-20dBm\\
Conversion efficiency for power transfer&30$\%$ \\
Battery capacity&37.7J \\
\hline
Power consumption in sleep mode& 1$\mu\text{W}$\\
\hline
\hline
\multicolumn{2}{|c|}{User sensor (low power): \textbf{accelerometer in pedometer application}}\\
\hline
Power consumption during active sensing & 28.5$\mu\text{W}$\\
Active sense time/Sensing period   & 1s/1s  \\
\hline 
\multicolumn{2}{|c|}{Operator sensor: \textbf{gas and volatile organic compounds sensor}}\\
\hline
Power consumption during active sensing & 32$\text{mW}$\\
Active sense time/Sensing period   & 25ms/100s \\
\hline
\hline
Payload, $D_k$   & 32B\\ 
\hline
\multicolumn{2}{|c|}{Radio option 1: \textbf{Bluetooth Smart in advertisement channels}}\\
\hline 
Transmission time/Report period   & $\left(0.15+\frac{D_k+10}{125}\right)$ms / 5s \\   
Consumed power&18.3mW\\ 
\hline
\hline
\multicolumn{2}{|c|}{Radio option 2: \textbf{LoRaWAN at its highest rate}}\\
\hline 
Transmission time/Report period   &  $\left(0.215+\frac{D_k+23}{6.25}\right)$ms / 5s\\      
Consumer power&40mW\\ 
\hline
\hline
\multicolumn{2}{|c|}{Radio option 3: \textbf{IEEE 802.15.4 2450 DSSS PHY}}\\
\hline
Transmission time/Report period   & $\left(1+\frac{D_k+15}{31.25}\right)$ms / 5s\\
Consumed power&18.3mW\\ 
\hline
\end{tabular}}
\vspace{-0.7cm}
\end{table}

\vspace{-0.3cm}
\section{System-Level Feasibility Study of WPCS}

In this section, we conduct a careful system-level feasibility study of our proposed WPCS system.

\vspace{-0.3cm}
\subsection{Considered Urban Modeling Setups}

\begin{figure}[!ht]
\centering
\includegraphics[width=1\columnwidth]{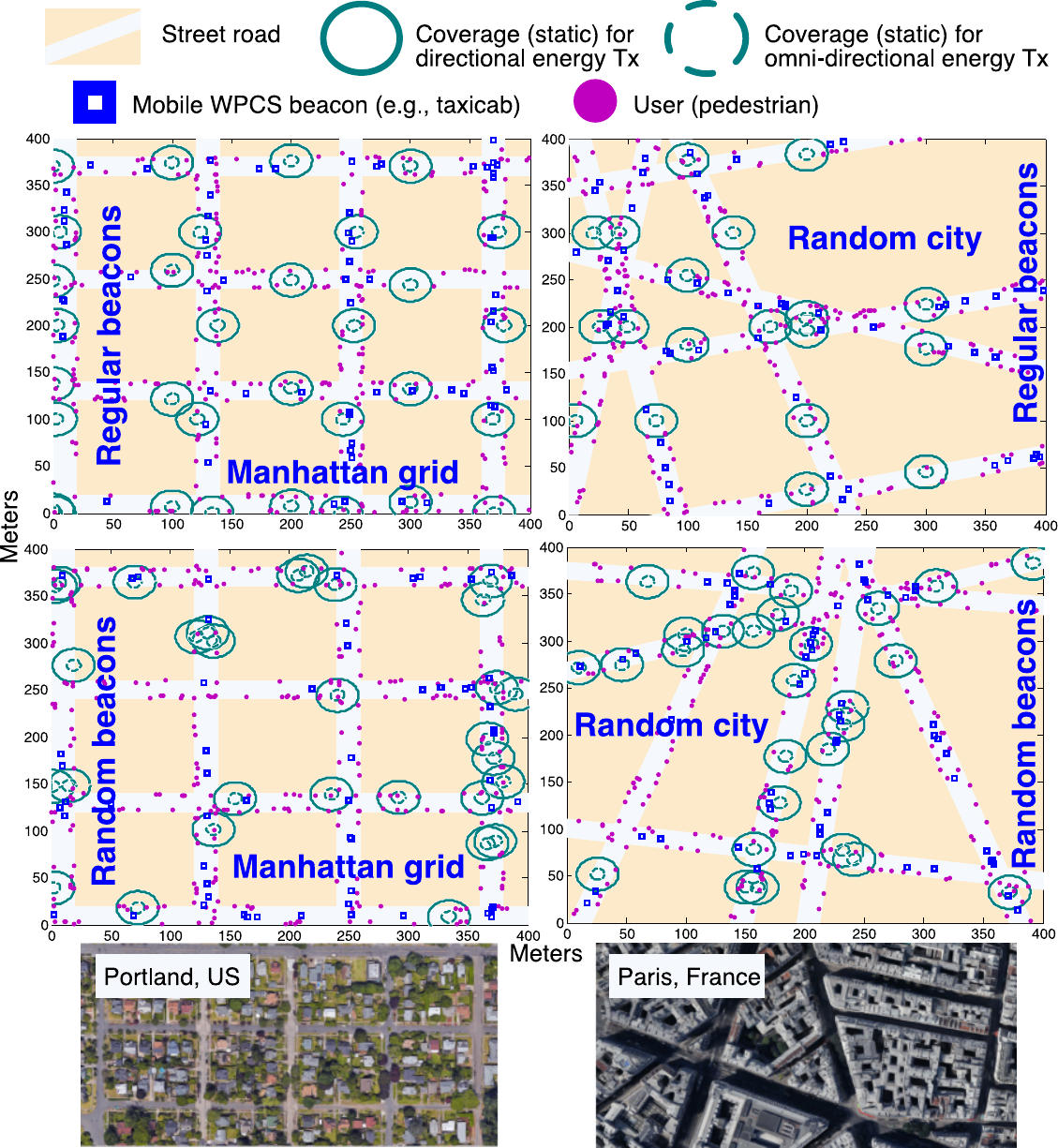}
\caption{Considered deployment types: Manhattan vs. random}
\label{fig:3}
\vspace{-0.7cm}
\end{figure}

We investigate the WPCS system behavior within an area of interest that represents a square with the side of $400$m (see Fig.~\ref{fig:3}). As our characteristic urban scenarios, we select (i) the Manhattan grid model (named here `Manhattan'), and (ii) a city layout of irregular structure (named here `random'). The latter is reconstructed multiple times throughout a simulation run, which allows to abstract away the particularities of individual instances thus arriving at the averaged characterization. The mean size of a city block is $100$x$100$m, whereas the street width is $20$m, of which the road occupies $5$m and the rest is pedestrian zone. The speed of vehicles is assumed to be $30$kmph, while people move at the speed that is distributed uniformly over $[3,6]$kmph. All mobile entities in our simulations move along the streets in their preferred direction and may turn left/right at intersections with equal probabilities. 

For the described city layouts (see Fig.~\ref{fig:3}), we assume that an operator may distribute WPCS beacons along the streets either regularly at a certain distance (named here `regular'), or randomly (named here `random'). Furthermore, every power beacon may either be stationary (located e.g., on lampposts and street furniture), or mobile (deployed e.g., on top of taxicabs). For each of our scenarios, there are two radio transmission modes: (i) all of the WPCS beacons exploit omnidirectional antennas, and (ii) directional antennas are utilized with no more than 6 simultaneous beams (see more details on directional WPT in~\cite{7462482}). When within coverage of several WPCS beacons, user's wearables are charged by all of them if not blocked physically. Here, the LoS blockage probability is calculated according to geometrical considerations by assuming e.g., that the current number of crowd sensing participants constitutes around $10\%$ of the total population.

We additionally differentiate between various types of wearables involved into different crowd sensing applications: users carry sensors for both \textit{personal} and \textit{collective} use (the sensors within a single wearable device are replaceable/removable). For a better representation of our below results, we impose that the battery is shared by e.g., one `personal' and one `collective' sensor (the latter is used for the crowd sensing application). We also take into account the situation when there is no crowd sensing payload involved and thus no wireless charging is offered (termed `default' in the figures). The rest of the system considerations are given in Table~\ref{sim_parameters}.

\vspace{-0.3cm}
\subsection{Representative Numerical Results}

\begin{figure*}
\centering
\includegraphics[width=2\columnwidth]{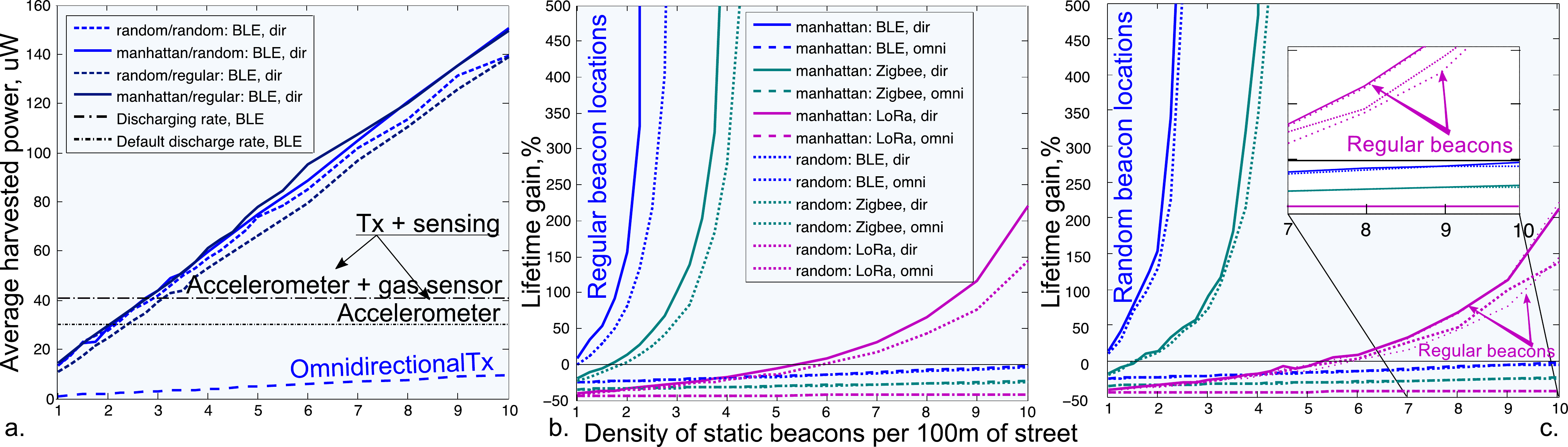}
\caption{User-centric performance perspective: amounts of harvested power averaged across users and time (a); gains in the average wearable device lifetime for regular (b) and random (c) static beacon locations vs. varying density of WPCS beacons}
\label{fig:4}
\vspace{-0.7cm}
\end{figure*}

Understanding the potential benefits for the user, we first evaluate the amounts of harvested power vs. the varying density of static WPCS beacons (see Fig.~\ref{fig:4}.a). We contrast the consumed power for the entire wearable device (both sensing and transmission components add up to the `discharge rate') in the default scenario (one personal sensor) against that in the WPCS scenario (personal and collective sensors). Along these lines, we initially focus on directional wireless charging (solid and dotted light/dark blue curves for all 4 scenarios). It can be observed that directional WPT remains sufficient to support the collective sensor itself. Moreover, starting at a certain beacon density, the received energy may guarantee sustainable operation of the entire wearable network based on Bluetooth low energy (BLE) radio technology.  
 
We also consider gains in wearable device lifetime (relative to the operation time in our default setup) by altering the radio technology (BLE, Zigbee, and LoRa). We investigate separately the regular and random static beacon deployments (see Fig.~\ref{fig:4}.b and~c) and learn that BLE outperforms other solutions, thus providing sustainable device operation at lower densities (solid blue, green, and purple curves). In contrast, Zigbee requires more beacons to reach the same effect, whereas LoRa cannot support sustainable operation due to its wider coverage. The above is only possible for directional charging, while omnidirectional WPT hardly compensates for the discharge rate. However, if WPT conversion efficiency grows beyond the assumed $30$\% in the future, omnidirectional charging may become preferred as it does not require complex positioning and beam steering.

\begin{figure}[!ht]
\vspace{-0.3cm}
\centering
\includegraphics[width=0.7\columnwidth]{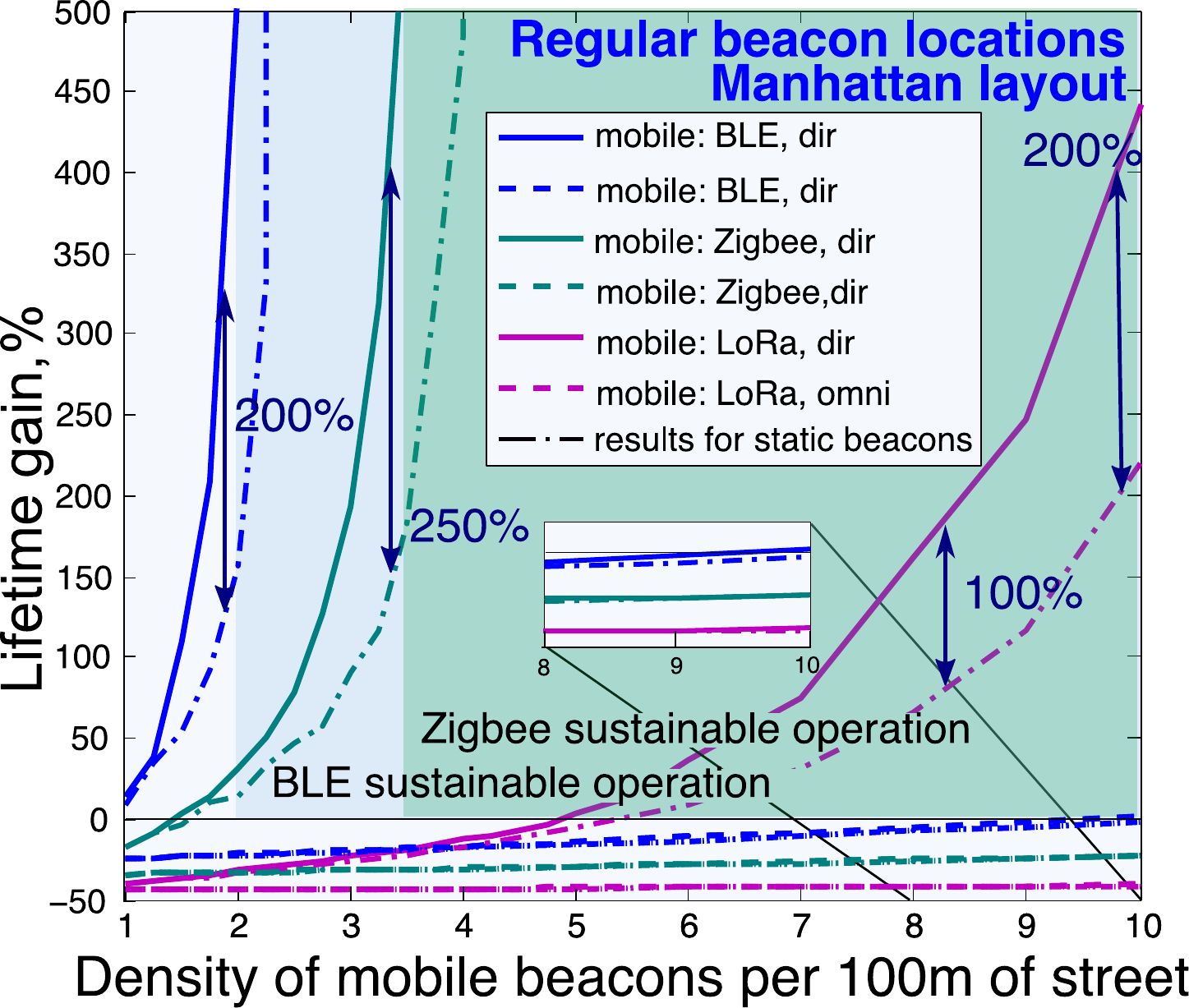}
\caption{Device lifetime evolution for mobile WPCS beacons}
\label{fig:5}
\vspace{-0.3cm}
\end{figure}

Comparing Fig.~\ref{fig:4}.b and~c, we also learn that our results are sensitive to the deployment type. Accordingly, the Manhattan scenario offers slightly better lifetime and sustainability performance, while being less sensitive to the WPCS beacon layout. By contrast, the random deployment prefers random beacon layout to regular, which poses an important question of optimized WPCS system planning for a particular urban landscape. We further select the Manhattan layout to demonstrate our WPCS system behavior in the presence of mobile beacons mounted on top of cars (the number of static beacons is set to zero here), as well as compare this case to the static results above. Interestingly, we may observe a significant difference in the device lifetime gains (see Fig.~\ref{fig:5}), which implies that under the same energy densities the mobile WPCS beacon deployment is more efficient.

Indeed, since the battery charge is constrained by its maximum capacity, the residence time within the beacon coverage plays a crucial role in providing operational sustainability, and the mobility of beacons naturally decreases the residence time as well as the time without the energy charge. This effect is clearly visible for directional WPT due to its wider radius, while the omnidirectional charging cannot enjoy similar performance. We also note that using more power-hungry wearables (sensors) may lead to somewhat less attractive results, thus yielding insignificant lifetime decrease due to participation in crowd sensing (e.g., around 1$\%$ for a heart rate monitor), while directional WPT technology enables only modest lifetime increase (up to $10$\%).

Exploring our WPCS system further, we additionally quantify a share of data obtained by the network operator when relying on two simple but rational user policies: (i) gathering and transmitting the sensed data only after the energy needed for the collective sensor is accumulated, and (ii) activating only if the harvested energy is sufficient for both the personal and the collective sensors. Then, for directional WPT the system operator obtains, respectively, $85$\% and $35$\% of information at the lowest WPCS beacon density, while collecting more at higher densities. For omnidirectional charging, the data acquisition remains from under $10$\% to $60$\% for policy one as well as $20$\% at most for policy two. Our presented results may open an important discussion on the human response and involvement, the beneficial user strategies, as well as the consequences of the former for the WPCS operator.

\vspace{-0.3cm}
\section{Prospective Energy--Data Trading Mechanisms}

In this section, we conclude our work by discussing the appropriate incentivization schemes for WPCS applications. 

\vspace{-0.3cm}
\subsection{Utilizing Auction Models for WPCS Interactions}

Realistically, the number of participants in the proposed WPCS system -- which can be represented as energy--data market -- may vary significantly subject to the actual interaction and incentivization mechanisms~\cite{7065282} adopted by the key stakeholders (operator(s), users, etc.). To this end, Auction Theory offers a set of powerful tools to design and optimize such mechanisms. Being an applied branch of Economics, it covers a wide range of trading and negotiation processes as well as delivers efficient rules and equilibrium strategies.

In current literature, the conventional and well-known types of auctions are: `English' (open ascending price), `Dutch' (open descending price), first-price sealed-bid, and second-price sealed-bid. Notably, all of these employ the \textit{winner-pays} rule, so that only the auction winner is bound to cover the agreed price. Therefore, for crowd sensing scenarios there were many attempts to alter the classical formulations with the \textit{all-pay} rule~\cite{Luo2016}, where all of the participants are required to commit their sensed information in advance -- hence effectively `paying' their bids regardless of the ultimate outcome. Even though all-pay auctions might seem to return higher profits, in practice however they see difficulty in incentivizing the participants to actually start bidding. Another downside of all-pay schemes is in their vulnerability to collusion between agents. A practical alternative to the all-pay model is to add the `lottery' flavor: offer every participant a non-zero probability to win where the chances are proportional to the bid size.

Both of the above considerations have recently received significant research attention and were thoroughly investigated within the context of mobile crowd sensing. However, for the battery constrained wearable devices, these generic approaches may eventually become limited, since excessive use of `collective' sensors may significantly decrease their lifetime (as we have seen in the previous section) as well as lower the motivation to participate for (risk-averse) users. In light of this, it might after all be helpful to consider the winner-pays auctions with the emphasis on \textit{divisible goods} (in our case, energy). We envisage that energy--data trading in WPCS systems may be modeled as (i) a \textit{multi-unit} auction with a \textit{single-unit} demand, when the network operator `sells' exactly $N$ positions for charging, or (ii) a \textit{share-auction}, when several winners share the energy resource (as they actually have to do anyway in the time domain, since the number of simultaneous WPT beams is inherently limited). 

In case of several winners, it is important to determine an effective selling strategy: (i) a \textit{uniform} pricing, when the winners pay the lowest win-price, or (ii) a \textit{discriminatory} pricing (e.g., `pay-as-bid'). Here, we argue that the \textit{sealed-bid} (as opposed to an \textit{open} auction) is not the only viable option -- although users have no explicit information about their competitors, it might become available implicitly through the online application data. Extending this formulation for the case of several WPCS system operators on the market may in turn require the consideration of \textit{reverse auctions}, where the agents are allowed to select a seller; or, in case of multiple agents, multiple sellers may establish e.g., an oligopoly game on the differentiated market. Further, automation of auction rules via cloud-based crowd sensing applications might then entail \textit{proxy-bidding} (as in case of eBay), when the system trades on behalf of the agent within adequate preset constraints.

\vspace{-0.3cm}
\subsection{Important Concluding Remarks and Future Work}

Applying auction schemes to design the WPCS-specific interaction rules leads not only to challenges typical for urban crowd sensing ecosystem (including incomplete and asymmetric information as well as stochastic population), but also to issues connected with the preferred mechanisms of trading. Given that battery charge plays the central role in this context, an intuitive factor for mapping user's value onto a respective bid may be rooted in how much the battery is discharged (i.e., a function of the discharge rate). Here, the availability of multiple candidate IoT technologies and wearable devices with their specific power consumption profiles creates unprecedented heterogeneity in the agent types as well as in the types of (probabilistic) knowledge about them. 

For the sake of modeling simplicity, a popular assumption -- same probabilistic properties for everyone -- may be adopted as the first step. Going further, additional research questions emerge that underpin the design of appropriate utility (payoff) functions for all the involved stakeholders, the definition of optimal online/local bidding schemes, the search for viable equilibrium strategies, the understanding of the conditions for collusion, as well as the improvements in auction design efficiency and revenue generation methods. Ultimately, the outlined research vectors will also include the need for comparative and quantitative description of how the potential trading/incentivization mechanisms impact the behavior of WPCS participants, the payoff of the network operator, the respective equilibrium points, and the overall social welfare levels. Addressing these important new challenges requires prompt attention of our entire research community.

\vspace{-0.3cm}
\bibliographystyle{ieeetr}
\bibliography{refs}

\end{document}